\pdfoutput=1
\documentclass[sigconf,screen=true,natbib=true]{acmart}

\usepackage{multirow}
\usepackage{enumitem}
\usepackage{stfloats}
\usepackage{pifont}
\usepackage{colortbl}
\definecolor{TabHL}{HTML}{DFE9F4}

\definecolor{MckBlue}{HTML}{005EB8}
\AtBeginDocument{\hypersetup{colorlinks=true, citecolor=MckBlue, linkcolor=MckBlue, urlcolor=MckBlue}}
\copyrightyear{2026}
\acmYear{2026}
\setcopyright{cc}
\setcctype{by}
\acmConference[ICAIF '26]{7th ACM International Conference on AI in Finance}{November 14--17, 2026}{Milan, Italy}
\acmBooktitle{7th ACM International Conference on AI in Finance (ICAIF '26), November 14--17, 2026, Milan, Italy}
\acmDOI{10.1145/3858063.3858909}
\acmISBN{979-8-4007-3087-0/2026/11}

\begin{document}

\title[LOB-Impact Benchmark]{LOB-Impact Benchmark: Emergent Market Impact in Generative Limit Order Book Models}

\author{George Nigmatulin}
\orcid{0000-0001-8935-1146}
\email{george.nigmatulin@eng.ox.ac.uk}
\affiliation{%
  \department{British Open-ended Learning and Discovery Lab (BOLD)}
  \department{Department of Engineering Science}
  \institution{University of Oxford}
  \city{Oxford}
  \country{United Kingdom}
}

\author{Valentin Mohl}
\orcid{0000-0002-6166-2222}
\email{valentin.mohl@cs.ox.ac.uk}
\affiliation{%
  \department{British Open-ended Learning and Discovery Lab (BOLD)}
  \department{Department of Computer Science}
  \institution{University of Oxford}
  \city{Oxford}
  \country{United Kingdom}
}

\author{Sascha Frey}
\orcid{0009-0001-6265-9607}
\email{sascha.frey@st-hughs.ox.ac.uk}
\affiliation{%
  \department{British Open-ended Learning and Discovery Lab (BOLD)}
  \department{Department of Computer Science}
  \institution{University of Oxford}
  \city{Oxford}
  \country{United Kingdom}
}

\author{Kang Li}
\orcid{0009-0007-7847-9292}
\email{kang.li@stats.ox.ac.uk}
\affiliation{%
  \department{British Open-ended Learning and Discovery Lab (BOLD)}
  \department{Department of Statistics}
  \institution{University of Oxford}
  \city{Oxford}
  \country{United Kingdom}
}

\author{Aramis Marti-Shahandeh}
\orcid{0009-0007-4757-4844}
\email{aramis@robots.ox.ac.uk}
\affiliation{%
  \department{British Open-ended Learning and Discovery Lab (BOLD)}
  \department{Department of Engineering Science}
  \institution{University of Oxford}
  \city{Oxford}
  \country{United Kingdom}
}

\author{Mihai Cucuringu}
\orcid{0000-0002-8464-2152}
\email{mihai@math.ucla.edu}
\affiliation{%
  \department{Department of Mathematics}
  \institution{University of California Los Angeles}
  \city{Los Angeles}
  \country{United States}
}
\affiliation{%
  \department{Department of Statistics \& Oxford Man Institute of Quantitative Finance}
  \institution{University of Oxford}
  \city{Oxford}
  \country{United Kingdom}
}

\author{Anisoara Calinescu}
\orcid{0000-0003-2082-734X}
\email{ani.calinescu@cs.ox.ac.uk}
\affiliation{%
  \department{British Open-ended Learning and Discovery Lab (BOLD)}
  \department{Department of Computer Science}
  \institution{University of Oxford}
  \city{Oxford}
  \country{United Kingdom}
}

\author{Jakob Foerster}
\orcid{0000-0001-9688-2498}
\email{jakob@robots.ox.ac.uk}
\affiliation{%
  \department{British Open-ended Learning and Discovery Lab (BOLD)}
  \department{Department of Engineering Science}
  \institution{University of Oxford}
  \city{Oxford}
  \country{United Kingdom}
}

\author{Stefan Zohren}
\orcid{0000-0002-3392-0394}
\email{stefan.zohren@eng.ox.ac.uk}
\affiliation{%
  \department{British Open-ended Learning and Discovery Lab (BOLD)}
  \department{Department of Engineering Science}
  \institution{University of Oxford}
  \city{Oxford}
  \country{United Kingdom}
}

\renewcommand{\shortauthors}{Nigmatulin, Mohl, Frey et al.}

\begin{abstract}
Realistic market impact is essential for generative limit order book (LOB) models to serve as world models for trading agents. If a simulator misprices this mechanism, agents learn policies that are optimal in the simulator but wrong in the real market. Existing metrics assess generated streams purely through distributional comparison with historical data, leaving counterfactual response to an agent's own orders largely unexplored. Here, we show that market impact emerges naturally in neural LOB generators trained solely for next-token prediction. We inject execution-style metaorders into the generated message flow of five autoregressive models across five NASDAQ stocks and measure the response using estimators from the empirical impact literature, validated on baselines whose physics is known in advance. Our benchmark evaluates any message-level generator in three phases: free generation (reading order-flow persistence against price diffusivity); metaorder injection (yielding the impact law's exponent and amplitude); and a post-execution window (measuring relaxation). In addition, comparing each visible injection to a blind, simulator-only control isolates the model's learned response from standard book mechanics. For every model on every stock, we measure how the impact responds as a function of cumulative metaorder participation. Interestingly, we observe emergent impact in generative models of market data, with exponents in the range of $0.63$ to $1.21$, slightly above the traditional $0.5$ of the square-root law of impact. In the family of models tested, we are so far unable to observe impact decay during relaxation. We tentatively attribute this to the still-short context windows of the models. We release the benchmark as open source to validate future generative models for market data.

\end{abstract}

\ccsdesc[500]{Computing methodologies~Simulation evaluation}
\ccsdesc[300]{Applied computing~Economics}

\keywords{limit order book, generative AI, market impact, square-root law, evaluation benchmark, state-space models, reinforcement learning}

\maketitle

\section{Introduction}
\label{sec:intro}

Generative LOB models (GANs~\citep{coletta2021towards}, recurrent and state-space message generators~\citep{nagy2023generative, smith2023s5, gu2023mamba, yang2024gated}, transformers and diffusion~\citep{li2024mars, backhouse2025painting}) are judged today by how closely their message streams resemble historical ones. Distributional benchmarks such as LOB-Bench~\citep{nagy2025lobbench} make this comparison precise by scoring spread, volume, inter-arrival, and imbalance distributions against real data. Such simulators are candidates to serve as world models for reinforcement-learning agents~\citep{mohl2025jaxmarlhft, frey2023jaxlob}, and an agent trained inside a simulator that gets the response to its orders wrong will learn mispriced policies, even if the surrounding stream looks realistic. Distributional scores are blind to this response, and existing benchmarks measure it only for a single event.

Constructing such a benchmark requires examining how real markets absorb large trades. Institutions do not execute a large order at once; they split it into small child orders. Empirical finance provides the reference for what such a metaorder does to the price: the square-root law of metaorder impact,
\begin{equation}
  I(Q) = \sigma \cdot Y \cdot \left(\tfrac{Q}{V}\right)^{\beta}, \qquad \beta \approx 0.5,
  \label{eq:sqrt_law}
\end{equation}
where $I(Q)$ is the expected price change, $\sigma$ the daily volatility, $Q$ the metaorder size, $V$ the daily volume, and $Y$ a dimensionless amplitude; the law is documented across equities, futures, and FX~\citep{almgren2005direct, toth2011anomalous, bouchaud2018trades}. This law is driven by two forces: \textbf{persistence} and \textbf{resilience}. Order flow is highly persistent: the sign of the next trade is predictable from the previous one~\citep{lillo2004long}. If every trade permanently moved the price, this persistence would cause indefinite trends. Instead, prices remain diffusive because the market absorbs and relaxes from each push~\citep{bouchaud2004fluctuations}. The law's concavity emerges when the liquidity a metaorder consumes is replenished behind it, and this same replenishment brings the price back down once execution stops~\citep{gatheral2010no}. In free generation, persistence is directly observable; resilience appears only indirectly as a diffusive price.

Isolating resilience therefore needs a counterfactual: how prices would have moved under an exogenous metaorder. A historical trace cannot answer that; a generator can. LOB-Bench measures this response at the scale of a single event. At the metaorder scale, MarS reports a square-root-shaped response to injected TWAP orders~\citep{li2024mars}, and \citet{bodor2025mdqr} an exponent of $0.55$ on a ten-child metaorder. However, these experiments do not separate the mechanical impact, where the injected orders remove levels of the price, from indirect impact through the reactions of other participants to those orders. In a generator, other participants are the model itself, and resilience lives in its reaction alone.

Our aim is to design experiments that measure these counterfactuals and isolate indirect impact. In our procedure, metaorders injected into five neural state-space generators produce an identifiable impact law on every stock we run: $25$ model--stock pairs, $25$ identifications. We provide an automatic per-stock calibration; every build-up injection is paired with a blind run that separates the model's reaction from the book's mechanics, and a number is reported only where a law is identified. This procedure extends distributional benchmarking, providing an instrument for evaluating the interactive fidelity of LOB simulators.

\begin{figure*}[t]
  \centering
  \includegraphics[width=\textwidth]{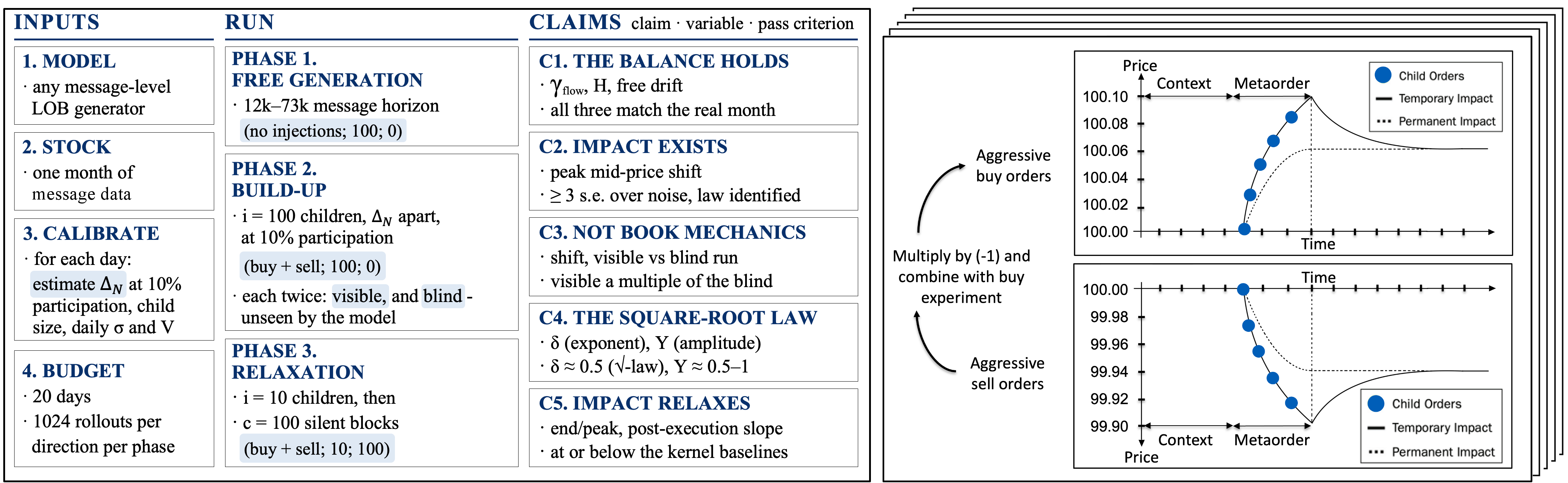}
  \Description{Two-part schematic. Left: a flow diagram of the benchmark, from four inputs through three phases (free generation, metaorder injection, post-execution) to five claims, C1 to C5, each with its variable and pass criterion. Right: timelines of the injection protocol, showing a context window, child orders injected at regular message intervals during execution, and a silent post-execution window, with the expected price path rising during execution and decaying afterwards.}
  \caption{\textbf{The benchmark and its counterfactual probe.} \textbf{Left:} four inputs, three phases, and the claim ladder they decide, each row reading \emph{claim $\cdot$ variable $\cdot$ pass criterion}. Phase~1 generates freely and decides the balance (C1); Phase~2 injects the build-up and decides existence (C2), attribution (C3) and the square-root law (C4); Phase~3 injects, then falls silent, and decides relaxation (C5). Schedules print as (what is injected; children $i$; silent blocks $c$). Every build-up injection runs twice, visible and \emph{blind}: same windows and seeds, the orders fill in the simulator while neither model input (message stream, book state) carries them, and the visible-over-blind ratio attributes the response to the model itself (C3). Table~\ref{tab:scorecard} prints the resulting ladder for every model $\times$ stock. \textbf{Right:} the injection protocol. After a conditioning window (\emph{Context}), aggressive children are injected \ensuremath{\Delta_N}{} model-generated messages apart; the post-execution phase injects none. The traced paths are the real-market expectation the benchmark tests against: a transient response decaying towards a permanent level. The sell experiment mirrors the buy; the two combine by antisymmetrisation (Eq.~\ref{eq:combined}), and the experiment repeats over many windows (stacked frames).}
  \label{fig:framework}
\end{figure*}

\section{Methodology}
\label{sec:method}

\subsection{Benchmark design}
The benchmark takes four inputs: the model under test, a stock, a calibration computed from that stock's own data (Section~\ref{sec:data}), and a budget of $n$ rollouts per direction per phase, spread over every trading day of the month. It runs three phases with two paired controls (Figure~\ref{fig:framework}, left). Everything downstream is fixed before any model is run: the injection schedule, calibration and estimator are set by the stock and the protocol, never by the model under test.

\paragraph{Phase 1: free generation.} The model runs freely over the injection horizon (the length of one build-up metaorder: 12k--73k messages depending on the stock, Table~\ref{tab:selected}).
For this phase, we measure the following three observables. The first is a consistency pair: $\gamma_{\mathrm{flow}}$, the decay exponent of the trade-sign autocorrelation (smaller means longer memory), read against $H$, the Hurst exponent of the generated mid-price ($H = 0.5$ is diffusive). Persistent signs must be compensated by transient impact, so a generator that combines persistent flow with a super-diffusive price ($H > 0.5$) has built a free lunch into its own market~\citep{bouchaud2004fluctuations}. The second is the free drift a rollout accumulates with nothing injected. The third is the price response to a single trade, against the real stream. Phase~1 serves as a first filter: the replay passes it perfectly yet has no impact at all, so establishing impact requires an intervention.

\paragraph{Phases 2 \& 3: metaorder injection and relaxation.} The injection realises the counterfactual of Section~\ref{sec:intro}: the same conditioning windows, run again under an exogenous metaorder. After a conditioning window from the real stream, we alternate one aggressive child of median trade size with \ensuremath{\Delta_N}{} generated messages, the spacing that holds participation at $\eta = 10\%$ (Section~\ref{sec:data}) (Figure~\ref{fig:framework}, right). Two shapes probe the two faces of impact. The \emph{build-up} shape $(i{=}100,\,c{=}0)$ reads the impact exponent along a $100$-child metaorder, a typical real length (Section~\ref{sec:data}). The \emph{relaxation} shape $(i{=}10,\,c{=}100)$ executes a short metaorder and then stops injecting for $100$ silent blocks: ten execution durations, over which real impact falls to about two-thirds of its peak within one to two durations and to about one-half by the tenth~\citep{moro2009market, bucci2019slow}. Every configuration runs in both directions, and the two are combined as
\begin{equation}
  I \;=\; \tfrac{1}{2}\,\bigl(I_{\mathrm{buy}} - I_{\mathrm{sell}}\bigr),
  \label{eq:combined}
\end{equation}
which also handles \emph{free drift}: direction-independent drift cancels in Eq.~\ref{eq:combined} and is measured separately on the no-injection arm, entering the accounting exactly once.

\paragraph{Blind control.} An identified law does not say whose law it is. The replay kernels of Table~\ref{tab:models}, with no model inside, are identified on AMD at $\delta = 0.73$--$0.81$, inside the neural range, so the exponent cannot attribute the response, and a large amplitude could be
purely mechanical impact.
The build-up therefore runs twice on the same windows and seeds, once visible and once \emph{blind}. In the blind run, the orders fill in the simulator, but neither model input (message stream, book state) carries them, so the generated flow proceeds as if the metaorder never happened.
This allows us to single out the indirect impact.
The laws are fitted on the visible runs. The blind control covers only the build-up: Phase~3 reads a ratio within one visible run, which a floor ten times below the peak cannot change.

The horizon is set by the object, not by the model. Market impact is a property of the metaorder, the working unit of optimal execution~\citep{gueant2016financial, bacry2015market}, and only at that scale can an emergent impact law be demonstrated. A build-up rollout therefore runs about $40$k messages on AMD, far past the $500$-message training window of latter-day models (Table~\ref{tab:models}). The no-injection arm (free generation at this horizon, the second paired control) also tracks rollout degradation.

Sample sizes are part of the protocol. Rollouts buy \emph{detection}: resolving a peak $P$ at three standard errors takes $n^{*} = 9\,\sigma_1^{2}/P^{2}$ per direction, $\sigma_1$ from a pilot. Trading days buy \emph{precision}: days are the independent unit and rollouts cannot substitute for them. A cell is printed only when its peak exceeds three standard errors and its law is identified. One campaign per stock ($1{,}024$ rollouts per direction per phase over $20$ trading days) meets both requirements, with about $1.4\times$ the minimum that subsampling shows is needed. With fewer than five days, the interval is often \emph{narrower} than the full-sample one, yet contains the full-sample value only $40$--$68\%$ of the time: the estimate looks precise but is wrong.

\begin{figure}[t]
  \centering
  \includegraphics[width=\columnwidth]{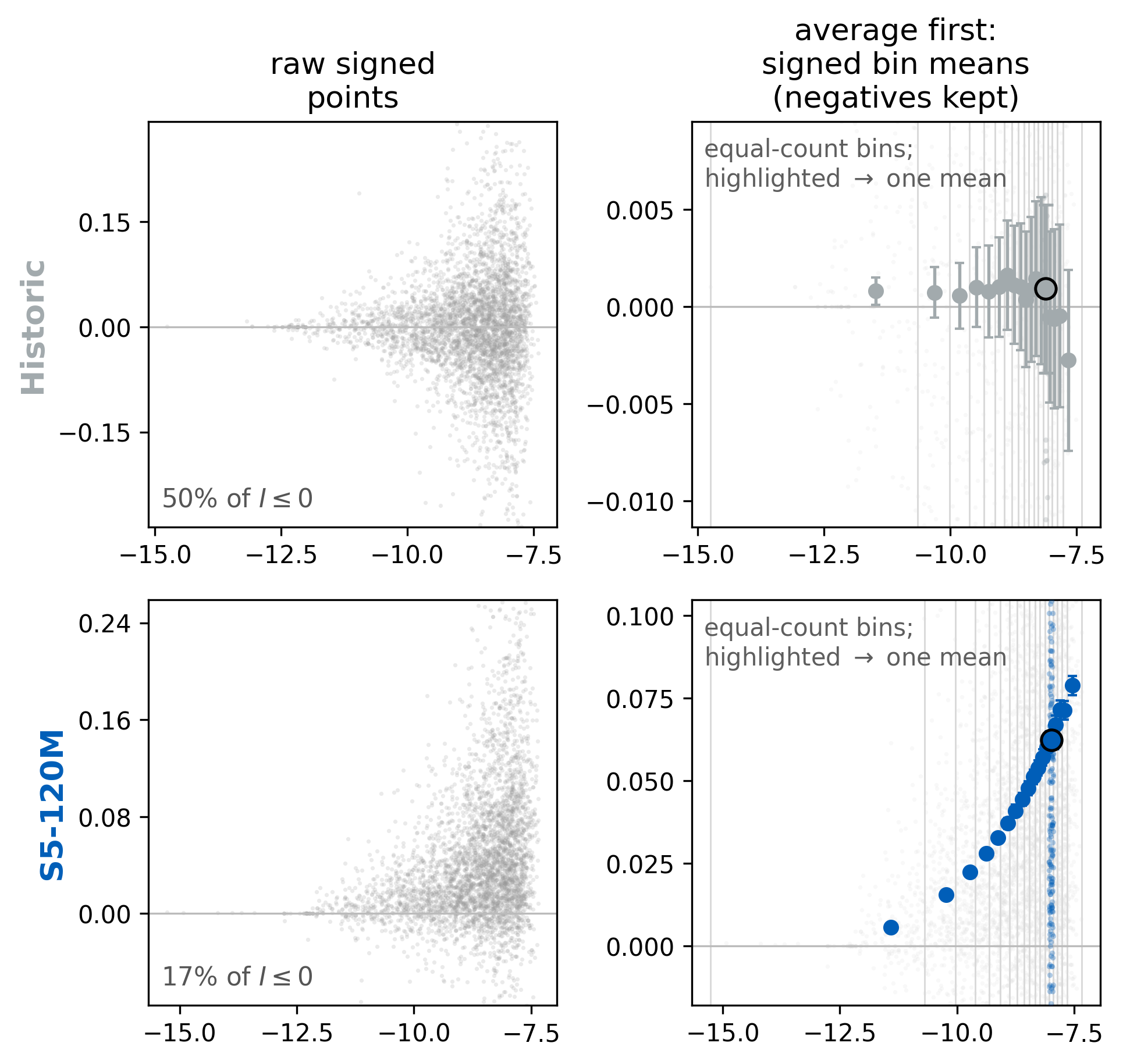}
  \Description{Four panels for AMD in a two-by-two grid; the top row is the historic replay and the bottom row is the S5-120M generator. The left panels show raw signed impact points against log participation. The right panels show equal-count bins with signed means, which straddle zero for the replay and rise for S5-120M.}
  \caption{\textbf{The estimator's input on AMD.} Rows: the Historic replay (negative control, top) and S5-120M (bottom). Columns: raw signed points $(\ln q, I)$ (left) and equal-count bins with \emph{signed} conditional means (right). Half the replay's points lie at or below zero; signed means cancel where there is no signal, so the replay's means straddle zero while S5-120M's rise. The estimator built on these means is defined in the text.}
  \label{fig:pipeline}\label{fig:loss}
\end{figure}

Each executed child $k$ contributes one signed point $(q_k, I_k)$, where $q_k = Q_k/V$ is the cumulative participation and $I_k$ the mid-price displacement. We write $\delta$ for the measured exponent and keep $\beta$ for the reference $0.5$ of Eq.~\ref{eq:sqrt_law}. The law $I = a\,q^{\delta}$ is fitted on the \emph{cumulative pool} $q \leq q_k$, because every prefix of a metaorder is itself a completed metaorder, the lifecycle convention of the empirical literature~\citep{bacry2015market, zarinelli2015beyond}. The local exponent declines along a metaorder, as for real ones~\citep{zarinelli2015beyond}, so $\delta$ is always quoted at the full metaorder, under one convention for every model.

Existence is decided before magnitude (Figure~\ref{fig:pipeline}). The identifiability test fits $I = a\,q^{\delta}$ directly by least squares, with no logarithms and no sign selection. For each $\delta$ on a grid ($[0.05, 1.5]$, step $0.01$) the best amplitude has a closed form, $a^{*}(\delta) = \sum I\,q^{\delta}/\sum q^{2\delta}$, so the misfit becomes a curve over $\delta$: a real law gives a sharp minimum, and no signal gives a flat curve with $a \approx 0$. A law counts as not identified when the amplitude is below the floor ($|Y| < 0.05$) or the best $\delta$ sits on a grid bound; the master curve (all sizes collapsed onto one) is drawn only when $\langle I(1)\rangle$ exceeds three standard errors. Testing existence first is the one step we add to an otherwise standard pipeline; a weighted-median $L_1$ version guards against outliers. The \emph{number}, quoted only where a law is identified, is the binned estimator of the empirical literature~\citep{sato2025universality}: $18$ quantile bins, \emph{signed} conditional bin means, OLS through the positive means.

All fits run on raw signed impact; volatility enters only as the \emph{unit} of the amplitude ($I$ in units of the day's Parkinson volatility), $Y = a\,q_{\mathrm{ref}}^{\,\delta-1/2}/\kappa_P$ at the data edge, with $\kappa_P = \sqrt{4\ln 2} \approx 1.665$ converting the Parkinson range~\citep{parkinson1980extreme} to the convention of the empirical studies, so $Y$ compares directly with the measured $Y \approx 0.5$~\citep{toth2011anomalous, zarinelli2015beyond, bucci2019crossover}. On AMD the binned $\delta$ moves by at most $0.05$ across six volatility normalisations (none, i.e.\ $\sigma = 1$, and the Parkinson, Garman--Klass, Rogers--Satchell, close-to-close and Yang--Zhang estimators), and the $L_1$ twin agrees wherever a law is identified.

Phase~3 reads relaxation through two numbers: the level at which impact settles and its slope after execution. We compare the level with real metaorders, which relax to one-half to two-thirds of their peak~\citep{moro2009market, bucci2019slow}, and with the kernel baselines run under the same schedule, whose closed-form value sets the scale of relaxation at this spacing. Fitted directly, the neural decay kernel comes out permanent ($\gamma \approx 0$) by two independent methods, schedule ordering and an impulse fit. Days are pooled on a common clock, the executed fraction $v = k/N$ of the metaorder, because metaorder impact does not depend on execution duration~\citep{bouchaud2018trades}; cells where fast and slow days still differ materially are marked $^{m}$.

The estimator is validated before it is read: laws of known exponent, planted in the replay's own point cloud (real noise, real day structure), are recovered across $\delta \in [0.5, 1.0]$ with a small downward bias (at most $0.11$ at neural amplitudes), so the exponents reported below are conservative. In $500$ null trials with nothing planted, the identifiability test returns not one false law. The kernel baselines close the loop end to end: Phase~3 recovers their built-in resilience in the correct order on all five stocks.

Each cell is finally reduced to a verdict on the five claims of Figure~\ref{fig:framework} (the C1--C5 column of Table~\ref{tab:scorecard}), on criteria fixed for every model: \textbf{C1} when $\gamma_{\mathrm{flow}}$, $H$ and the free drift sit within tolerances fixed in advance ($0.10$, $0.02$ and $5$ ticks) of the stock's own replay; \textbf{C2} when the gate identifies a law; \textbf{C3} when the visible peak is at least $3\times$ the blind floor; \textbf{C4} when $|\delta - 0.5| \leq 0.10$ and $Y \in [0.5, 1]$; \textbf{C5} when the end-over-peak is at or below that of the Obizhaeva--Wang (OW) kernel, the standard resilient book, on the same stock and schedule, with a non-positive slope and a free drift under $20$ ticks, so the descent is not the drift; the OW row itself is left undecided. C4 and C5 are undecidable when no law is identified, and C3 when no blind run was made, so a missing measurement cannot read as a defect. Every verdict follows from the numbers printed beside it, so the whole ladder can be checked from Table~\ref{tab:scorecard} alone.

\vfill
\subsection{Models under test}
\label{sec:models}

\begin{table*}[t!]
  \centering
  \caption{\textbf{Models under test.} Top block: non-neural baselines, calibrated (where anything is fitted) on the evaluation stock itself. Bottom block: neural generators sharing one corpus (merged cell) that \emph{includes} AMD, so every neural measurement is in-distribution. ``Gen.''\ is the wall-clock for one build-up rollout (${\sim}40$k messages): replay baselines transform a stored stream (---), parametric baselines run on CPU, neural models on one H100. $^{\dagger}$Fine-tuned with a 2000-message context and run with a 4000-message window.}
  \label{tab:models}
  \footnotesize\renewcommand{\arraystretch}{0.92}
  \begin{tabular}{@{}llp{0.36\textwidth}p{0.22\textwidth}r@{}}
    \toprule
    Model & Class & How it works & Trained / calibrated on & Gen. \\
    \midrule
    Historic    & replay            & Replays the real stream unchanged; cannot react. & real held-out data & --- \\
    Heuristic   & replay $+$ shift  & Replay; a consumed best level shifts all later messages. & rule fixed a priori & --- \\
    Propagator  & replay $+$ kernel & Replay $+$ one-tick kicks decaying as $\tfrac{2}{3}+\tfrac{1}{3}\ell^{-1/2}$. & kernel from theory~\citep{bouchaud2018trades} & --- \\
    OW          & replay $+$ kernel & Replay $+$ Obizhaeva--Wang exponential-resilience kernel. & kernel from theory~\citep{obizhaeva2013optimal} & --- \\
    CST         & parametric        & Memoryless event rates vs distance from mid~\citep{cont2010stochastic}. & rates on the stock, Jan 2026 & ${\sim}4$\,s \\
    NMZI        & parametric        & CST $+$ return-driven order-sign feedback~\citep{ravagnani2026nmzi}. & as CST; feedback at defaults & ${\sim}4$\,s \\
    Hawkes      & point process     & Self-/cross-exciting intensities; short-range clustering~\citep{bacry2015hawkes}. & kernels on the stock, Jan 2026 & ${\sim}4$\,s \\
    QR          & parametric        & Queue-reactive intensities conditioned on the book~\citep{huang2015queue}. & intensities on the stock, Jan 2026 & ${\sim}4$\,s \\
    \midrule
    S5-120M     & neural SSM        & Autoregressive next-token generation; S5 backbone~\citep{smith2023s5}. & \multirow{5}{=}{\raggedright GOOG, AAPL, NVDA, AMZN, META, TSLA, MSFT, \textbf{AMD}; Jan 2022--Dec 2025; Jan 2026 (the evaluation month) held out.} & $53$\,s \\
    S5-4k       & neural SSM        & S5 with a 4000-message conditioning window. & & ${\sim}12$\,min \\
    Mamba3      & neural SSM        & Selective state-space backbone~\citep{gu2023mamba, lahoti2026mamba3}. & & $39$\,s \\
    Mamba3-4k   & neural SSM        & Mamba-3 with a 4000-message conditioning window$^{\dagger}$. & & ${\sim}12$\,min \\
    GDN         & neural SSM        & Gated-DeltaNet recurrence~\citep{yang2024gated}. & & $54$\,s \\
    \bottomrule
  \end{tabular}
\end{table*}

\begin{figure*}[t!]
  \centering
  \includegraphics[width=.95\textwidth]{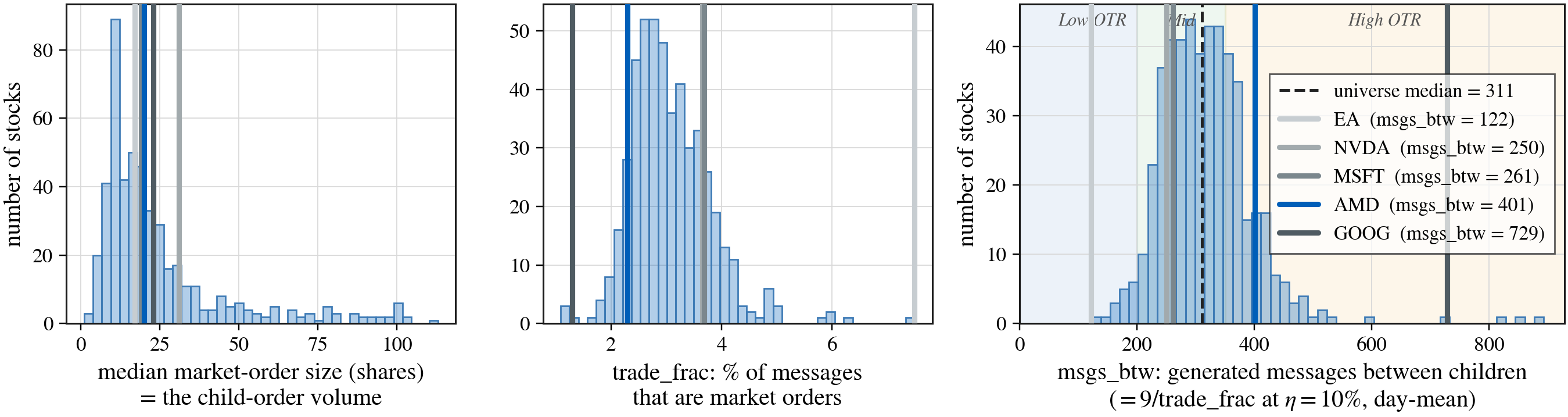}
  \Description{Three histograms over 487 stocks: the median market-order size, the share of messages that are market orders, and the number of generated messages between child orders. Shaded bands mark OTR cohorts, and vertical lines mark the universe median and the five selected stocks.}
  {\captionsetup{type=figure}\caption{\textbf{The universe and the calibration it fixes} (day-mean over $487$ names, January 2026). \textbf{Left:} the median market-order size, which sets the child volume. \textbf{Middle:} the market-order fraction of the message stream, the denominator of Eq.~\ref{eq:mb}. \textbf{Right:} the resulting \ensuremath{\Delta_N}{}, the messages the generator must produce between children at $\eta = 10\%$; shaded bands are the OTR cohorts, vertical lines the universe median and the five selected stocks. Since \ensuremath{\Delta_N}{} is the generation burden of the \emph{same} experiment, the five span the universe minimum (EA), the modal band (NVDA, MSFT), the high-OTR cohort (AMD) and its tail (GOOG).}
  \label{fig:sp500hist}}
\end{figure*}

The benchmark accepts any message-level generator; Table~\ref{tab:models} lists the panel of models we test. The replay block spans no causal response (Historic, the negative control), permanent impact (Heuristic), and two kernels with the target physics written in: the Propagator, whose power-law decay to $\tfrac{2}{3}$ is the positive control, and Obizhaeva--Wang (OW), with exponential resilience. The five neural generators carry a continuous hidden state across generated and injected messages, the mechanism through which relaxation could arise, and share one eight-name corpus including AMD, evaluation month held out. Book reconstruction and matching use the JAX-LOB engine~\citep{frey2023jaxlob}; every number comes from the one campaign per stock of Section~\ref{sec:method}.

\subsection{Data preparation}
\label{sec:data}

The injection protocol requires three calibration numbers per stock: the child size, the spacing between children, and the daily volume that normalises cumulative participation. All three are computed from LOBSTER message data for January 2026, the evaluation month, held out of model training.

As impact is measured relative to the stock's own activity, the measurements are re-calibrated per name. We choose a  \emph{participation target} of $\eta = 10\%$ of the market orders during execution; broker databases retain execution participations of $3$--$40\%$~\citep{bacry2015market} and the empirical impact surface is fitted over $10^{-4}$ to $0.1$~\citep{zarinelli2015beyond}. The child volume is the stock's median market-order size, $V$ is the day's volume, and the spacing follows from the participation identity $\eta = 1/(1 + \ensuremath{\Delta_N}\,\phi)$:
\begin{equation}
  \ensuremath{\Delta_N}\;=\; \frac{\eta^{-1} - 1}{\phi} \;=\; \frac{9}{\phi}\,,
  \label{eq:mb}
\end{equation}
computed per day from one month of LOBSTER message data~\citep{lobster2023}. Here $\phi$ is the market-order share of the day's message stream (Figure~\ref{fig:sp500hist}, middle). The \emph{child count} $N = 100$ is fixed a priori, never per model: the mean metaorder length in the empirical samples ($96$--$98$ trades per hidden order~\citep{moro2009market}, $73$--$100$ child orders~\citep{bacry2015market}), spanning two decades of cumulative participation. A shorter metaorder would sample only the early, steepest part and bias $\delta$ upward.

Figure~\ref{fig:sp500hist} shows the universe: \ensuremath{\Delta_N}{} is strongly right-skewed, so a high-OTR name (order-to-trade ratio $1/\phi$) costs seven times more generation than a low-OTR one. Table~\ref{tab:selected} takes one name per cohort. This paper reports AMD band by band: high-OTR cohort ($\ensuremath{\Delta_N}= 401 \pm 69$, ${\approx}40$k messages per metaorder), and inside every training corpus, so the measurement carries no distribution shift.

\begin{table}[H]
  \caption{\textbf{Selected stocks, one per cohort} (day-mean, whole shares, January 2026). ``Child'' is the median market-order size, used as the child size; $\pm$ = std across days; ``gen'' is the messages a $100$-child build-up metaorder must generate. Figures~\ref{fig:pipeline} and~\ref{fig:flowbal} show AMD; the others are in Table~\ref{tab:scorecard} only.}
  \label{tab:selected}
  \centering\footnotesize
  \begin{tabular}{@{}llcccc@{}}
    \toprule
    Stock & Cohort & Child ($\pm$std) & \ensuremath{\Delta_N}{} ($\pm$std) & $\phi$ & gen \\
    \midrule
    EA   & Low OTR & $17\,(\pm4)$ & $122\,(\pm15)$ & $7.49\%$ & $12$k \\
    NVDA & Mid (modal) & $31\,(\pm7)$ & $250\,(\pm35)$ & $3.65\%$ & $25$k \\
    MSFT & Mid (modal) & $19\,(\pm3)$ & $261\,(\pm59)$ & $3.68\%$ & $26$k \\
    \textbf{AMD} & High OTR & $20\,(\pm3)$ & $401\,(\pm69)$ & $2.30\%$ & $40$k \\
    GOOG & High OTR (tail) & $23\,(\pm2)$ & $729\,(\pm183)$ & $1.30\%$ & $73$k \\
    \bottomrule
  \end{tabular}
\end{table}

\vfill\newpage
\section{Results}
\label{sec:results}

\begin{figure*}[p]
  \centering
  \includegraphics[width=\textwidth]{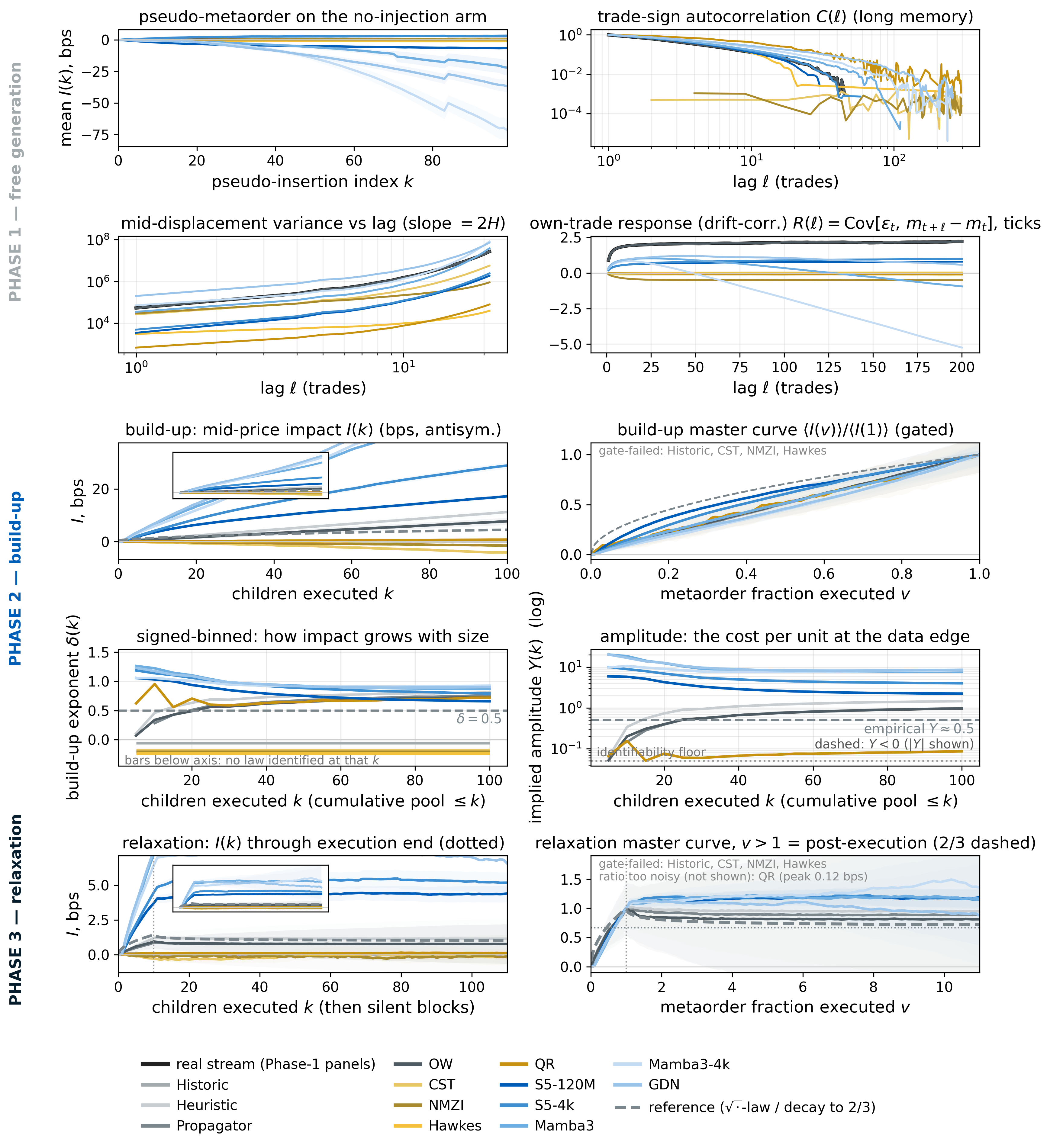}
  \Description{Ten panels for AMD in five rows of two. The top rows show Phase 1: pseudo-metaorder impact without injection, trade-sign autocorrelation, mid-price variance scaling and the own-trade response. The middle rows show Phase 2: impact against a square-root reference, a master curve, and binned estimates of exponent and amplitude along the metaorder. The bottom row shows Phase 3: impact after execution ends and the master curve against a theoretical decay reference.}
  \caption{The AMD measurements, all three phases in one figure ($5{\times}2$). \textbf{Top, Phase~1}: pseudo-metaorders on the no-injection arm (the pipeline returns drift, not impact); the trade-sign autocorrelation; the mid-displacement variance (slope $=2H$); the drift-corrected own-trade response $R(\ell)$. \textbf{Middle, Phase~2}: antisymmetrised $I(k)$ against the $\sqrt{\cdot}$-law reference (full range inset), the master curve, and the binned estimator along the metaorder ($\delta(k)$ and $Y(k)$, both against $0.5$). \textbf{Bottom, Phase~3}: $I(k)$ past the end of execution, and the master curve against the theoretical reference (dashed: $\sqrt{v}$ build-up, then power-law decay to $\tfrac{2}{3}$). Apart from GDN's drift, only the curves with resilience typed in come down.}
  \label{fig:flowbal}\label{fig:midtraj}\label{fig:beta3}\label{fig:relax}
\end{figure*}

Table~\ref{tab:scorecard} gives every measurement for the $13$ models on all five stocks. On every stock the neural generators produce an impact law of their own, but it is not the square-root law, and it does not relax. Below we go through the AMD measurements (Figure~\ref{fig:flowbal}) phase by phase; Section~\ref{sec:discussion} interprets them.

\subsection{Phase 1: Free generation}
\label{sec:res-flow}

In identical windows, the real AMD stream is balanced: persistent trade signs ($\gamma_{\mathrm{flow}} = 1.66$) with a diffusive mid-price ($H = 0.54$). No generator holds both legs. The Mamba family carries longer flow memory than the real stream ($\gamma_{\mathrm{flow}} = 1.20$ and $1.06$) and a price that is too diffusive ($H = 0.58$ for both): persistence left uncompensated. The S5 family carries shorter flow memory ($1.90$, $1.83$) with the real diffusivity ($H = 0.55$, $0.54$); GDN carries the panel's longest flow memory ($0.79$) with a price below diffusive ($H = 0.49$). Every neural generator also drifts when left alone, from $+9$ (S5-4k) to $-251$ (Mamba3-4k) ticks per rollout (Table~\ref{tab:scorecard}). Every neural generator therefore fails C1.

\begin{table*}[p]
  \captionsetup{font=footnotesize}
  \caption{All five stocks, one row per model, the same measurements everywhere (the AMD block is the one read band by band in Section~\ref{sec:results}). Columns: Par.\ = parameters; Ctx = conditioning window; LOB-Bench KS/L1/WS on free-generation rollouts against the real stream, 500-message windows, lower is better; Phase-1 flow balance ($\gamma_{\mathrm{fl}}$ = trade-sign memory, $H$ = diffusivity) and free drift (ticks); Phase-2 build-up ($\delta$ = binned exponent, where the direct-$L_2$ profile identifies one; $\pm$ = half-width of the 95\% day-clustered bootstrap interval, the trading day being the resampling unit; $Y$ = amplitude at the data edge; v/b = visible peak over blind floor, the C3 ratio); Phase-3 relaxation (e/p = end-over-peak; slp = post-execution slope in \% of peak per execution duration). \emph{target} rows: the stock's real stream in identical windows; (exp) = the empirical value for real markets, from the literature. C1--C5 = the claim ladder of Figure~\ref{fig:framework}, one verdict per claim (\ding{51}\,\ pass, \ding{55}\,\ fail, {\color{black!55}\raisebox{0.15ex}{\scalebox{1.35}{$\cdot$}}}\,\ not decidable) on the criteria of Section~\ref{sec:method}: C1 \emph{the balance holds}; C2 \emph{impact exists}; C3 \emph{not book mechanics}; C4 \emph{the square-root law}; C5 \emph{the impact relaxes}. The blind arm behind C3 runs for the neural panel only. \colorbox{TabHL}{Shaded bold}: the two numbers behind the neural panel's failed C4 and C5 -- $\delta$ against the $0.5$ of the square-root law, e/p against the real-market $\tfrac{2}{3}$ relaxation (the kernel rows bound this schedule) -- the one signature all $25$ neural cells repeat. Markers: ``---'' does not exist; n.i.\ = not identified; ($\cdot$)$^{u}$ = below the identifiability or sample-size requirement, shown for reference only; $^{c}$ by construction; $^{m}$ duration-mixed; $^{f}$ fixed a priori; $^{\dagger}$2k fine-tune at 4000; $^{\ddagger}$EA out-of-distribution.}
  \label{tab:scorecard}\label{tab:dashboard}
  \centering\scriptsize
  \renewcommand{\arraystretch}{0.65}
  \begin{tabular}{@{}l rr | ccc | ccc | ccc | cc | c@{}}
    \toprule
    Model & Par. & Ctx & KS & L1 & WS & $\gamma_{\mathrm{fl}}$ & $H$ & drift & $\delta$ & $Y$ & v/b & e/p & slp & C1--C5 \\
    \midrule
    \multicolumn{15}{@{}l}{\textbf{AMD}} \\
    \emph{target} & --- & --- & low & low & low & $1.66$ & $0.54$ & ${\approx}0$ & {\boldmath$0.5$}\,{\tiny(exp)} & $0.5$\,{\tiny(exp)} & --- & {\boldmath$\tfrac{2}{3}$}\,{\tiny(exp)} & ${\sim}{-1.5}$\,{\tiny(exp)} & --- \\
    \midrule
    Historic & --- & --- & .060 & .162 & .096 & $1.66$ & $0.54$ & $+1$ & ($0.11$)$^{u}$ & ($0.03$)$^{u}$ & --- & ($-0.00$)$^{u}$ & ($-0.5$)$^{u}$ & \ding{51}\,\ding{55}\,{\color{black!55}\raisebox{0.15ex}{\scalebox{1.35}{$\cdot$}}}\,{\color{black!55}\raisebox{0.15ex}{\scalebox{1.35}{$\cdot$}}}\,{\color{black!55}\raisebox{0.15ex}{\scalebox{1.35}{$\cdot$}}}\,\\
    Heuristic & 2$^{f}$ & $\infty$ & .060 & .162 & .096 & $1.66$ & $0.54$ & $+1$ & $0.81$ & $1.46$ & --- & $0.95$ & $-0.3^{c}$ & \ding{51}\,\ding{51}\,{\color{black!55}\raisebox{0.15ex}{\scalebox{1.35}{$\cdot$}}}\,\ding{55}\,\ding{55}\,\\
    Propagator & 2$^{f}$ & $\infty$ & .060 & .162 & .096 & $1.66$ & $0.54$ & $+1$ & $0.76$ & $0.98$ & --- & $0.89$ & $-0.5$ & \ding{51}\,\ding{51}\,{\color{black!55}\raisebox{0.15ex}{\scalebox{1.35}{$\cdot$}}}\,\ding{55}\,\ding{55}\,\\
    OW & 2$^{f}$ & exp. & .041 & .106 & .070 & $1.66$ & $0.54$ & $+1$ & $0.73$ & $0.96$ & --- & $0.81$ & $-0.3$ & \ding{51}\,\ding{51}\,{\color{black!55}\raisebox{0.15ex}{\scalebox{1.35}{$\cdot$}}}\,\ding{55}\,{\color{black!55}\raisebox{0.15ex}{\scalebox{1.35}{$\cdot$}}}\,\\
    CST & 1.5k & $0$ & .328 & .523 & .570 & n.i. & $0.47$ & $+1$ & n.i. & n.i. & --- & ($0.68$)$^{u}$ & ($-10.3$)$^{u}$ & {\color{black!55}\raisebox{0.15ex}{\scalebox{1.35}{$\cdot$}}}\,\ding{55}\,{\color{black!55}\raisebox{0.15ex}{\scalebox{1.35}{$\cdot$}}}\,{\color{black!55}\raisebox{0.15ex}{\scalebox{1.35}{$\cdot$}}}\,{\color{black!55}\raisebox{0.15ex}{\scalebox{1.35}{$\cdot$}}}\,\\
    NMZI & 1.5k$+$2 & EWMA & .330 & .526 & .574 & n.i. & $0.27$ & $0$ & n.i. & n.i. & --- & ($0.73$)$^{u}$ & ($+3.6$)$^{u}$ & {\color{black!55}\raisebox{0.15ex}{\scalebox{1.35}{$\cdot$}}}\,\ding{55}\,{\color{black!55}\raisebox{0.15ex}{\scalebox{1.35}{$\cdot$}}}\,{\color{black!55}\raisebox{0.15ex}{\scalebox{1.35}{$\cdot$}}}\,{\color{black!55}\raisebox{0.15ex}{\scalebox{1.35}{$\cdot$}}}\,\\
    Hawkes & 0.4k & ${\lesssim}1$s & .477 & .534 & .644 & $2.23$ & $0.20$ & $-3$ & ($0.48$)$^{u}$ & ($0.03$)$^{u}$ & --- & ($1.04$)$^{u}$ & ($-1.9$)$^{u}$ & \ding{55}\,\ding{55}\,{\color{black!55}\raisebox{0.15ex}{\scalebox{1.35}{$\cdot$}}}\,{\color{black!55}\raisebox{0.15ex}{\scalebox{1.35}{$\cdot$}}}\,{\color{black!55}\raisebox{0.15ex}{\scalebox{1.35}{$\cdot$}}}\,\\
    QR & 90 & $0$ & .445 & .503 & .607 & $.88$ & $0.42$ & $+0$ & $0.72$ & $0.09$ & --- & $1.12$ & $+2.6$ & \ding{55}\,\ding{51}\,{\color{black!55}\raisebox{0.15ex}{\scalebox{1.35}{$\cdot$}}}\,\ding{55}\,\ding{55}\,\\
    S5-120M & 120M & 500 & .142 & .231 & .185 & $1.90$ & $0.55$ & $-18$ & \cellcolor{TabHL}{\boldmath $0.66{\pm}.05$} & $2.24$ & $21$ & \cellcolor{TabHL}{\boldmath $1.19$} & $+0.6$ & \ding{55}\,\ding{51}\,\ding{51}\,\ding{55}\,\ding{55}\,\\
    S5-4k & 55M & 4k & .171 & .270 & .237 & $1.83$ & $0.54$ & $+9$ & \cellcolor{TabHL}{\boldmath $0.79{\pm}.06$} & $4.00$ & $26$ & \cellcolor{TabHL}{\boldmath $1.15$} & $+0.0$ & \ding{55}\,\ding{51}\,\ding{51}\,\ding{55}\,\ding{55}\,\\
    Mamba3 & 78M & 500 & .106 & .244 & .144 & $1.20$ & $0.58$ & $-69$ & \cellcolor{TabHL}{\boldmath $0.87{\pm}.06$} & $7.65$ & $9.5$ & \cellcolor{TabHL}{\boldmath $1.16^{m}$} & $+0.4$ & \ding{55}\,\ding{51}\,\ding{51}\,\ding{55}\,\ding{55}\,\\
    Mamba3-4k & 78M & 4k$^{\dagger}$ & .128 & .245 & .165 & $1.06$ & $0.58$ & $-251$ & \cellcolor{TabHL}{\boldmath $0.93{\pm}.05$} & $8.38$ & $5.0$ & \cellcolor{TabHL}{\boldmath $1.36^{m}$} & $+3.7$ & \ding{55}\,\ding{51}\,\ding{51}\,\ding{55}\,\ding{55}\,\\
    GDN & 94M & 500 & .139 & .250 & .183 & $.79$ & $0.49$ & $-127$ & \cellcolor{TabHL}{\boldmath $0.90{\pm}.05$} & $8.57$ & $12$ & \cellcolor{TabHL}{\boldmath $0.88^{m}$} & $-2.4$ & \ding{55}\,\ding{51}\,\ding{51}\,\ding{55}\,\ding{55}\,\\
    \midrule
    \multicolumn{15}{@{}l}{\textbf{NVDA}} \\
    \emph{target} & --- & --- & low & low & low & $1.31$ & $0.49$ & ${\approx}0$ & {\boldmath$0.5$}\,{\tiny(exp)} & $0.5$\,{\tiny(exp)} & --- & {\boldmath$\tfrac{2}{3}$}\,{\tiny(exp)} & ${\sim}{-1.5}$\,{\tiny(exp)} & --- \\
    \midrule
    Historic & --- & --- & .054 & .118 & .068 & $1.31$ & $0.49$ & $-1$ & ($-0.33$)$^{u}$ & ($0.00$)$^{u}$ & --- & ($0.00$)$^{u}$ & ($-0.3$)$^{u}$ & \ding{51}\,\ding{55}\,{\color{black!55}\raisebox{0.15ex}{\scalebox{1.35}{$\cdot$}}}\,{\color{black!55}\raisebox{0.15ex}{\scalebox{1.35}{$\cdot$}}}\,{\color{black!55}\raisebox{0.15ex}{\scalebox{1.35}{$\cdot$}}}\,\\
    Heuristic & 2$^{f}$ & $\infty$ & .054 & .118 & .068 & $1.31$ & $0.49$ & $-1$ & $0.93$ & $0.99$ & --- & $1.07$ & $+0.1^{c}$ & \ding{51}\,\ding{51}\,{\color{black!55}\raisebox{0.15ex}{\scalebox{1.35}{$\cdot$}}}\,\ding{55}\,\ding{55}\,\\
    Propagator & 2$^{f}$ & $\infty$ & .054 & .118 & .068 & $1.31$ & $0.49$ & $-1$ & $0.84$ & $0.66$ & --- & $1.04$ & $+0.1$ & \ding{51}\,\ding{51}\,{\color{black!55}\raisebox{0.15ex}{\scalebox{1.35}{$\cdot$}}}\,\ding{55}\,\ding{55}\,\\
    OW & 2$^{f}$ & exp. & .055 & .098 & .061 & $.97$ & $0.51$ & $+1$ & $0.83$ & $0.69$ & --- & $0.92$ & $+0.0$ & \ding{55}\,\ding{51}\,{\color{black!55}\raisebox{0.15ex}{\scalebox{1.35}{$\cdot$}}}\,\ding{55}\,{\color{black!55}\raisebox{0.15ex}{\scalebox{1.35}{$\cdot$}}}\,\\
    CST & 1.5k & $0$ & .219 & .398 & .303 & $.16$ & $0.44$ & $0$ & n.i. & n.i. & --- & ($0.64$)$^{u}$ & ($-9.2$)$^{u}$ & \ding{55}\,\ding{55}\,{\color{black!55}\raisebox{0.15ex}{\scalebox{1.35}{$\cdot$}}}\,{\color{black!55}\raisebox{0.15ex}{\scalebox{1.35}{$\cdot$}}}\,{\color{black!55}\raisebox{0.15ex}{\scalebox{1.35}{$\cdot$}}}\,\\
    NMZI & 1.5k$+$2 & EWMA & .224 & .399 & .308 & $-.12$ & $0.38$ & $0$ & n.i. & n.i. & --- & ($1.06$)$^{u}$ & ($-1.0$)$^{u}$ & \ding{55}\,\ding{55}\,{\color{black!55}\raisebox{0.15ex}{\scalebox{1.35}{$\cdot$}}}\,{\color{black!55}\raisebox{0.15ex}{\scalebox{1.35}{$\cdot$}}}\,{\color{black!55}\raisebox{0.15ex}{\scalebox{1.35}{$\cdot$}}}\,\\
    Hawkes & 0.4k & ${\lesssim}1$s & .259 & .339 & .345 & $1.71$ & $0.20$ & $+2$ & ($1.05$)$^{u}$ & ($0.65$)$^{u}$ & --- & ($-3.00$)$^{u}$ & ($-59.5$)$^{u}$ & \ding{55}\,\ding{55}\,{\color{black!55}\raisebox{0.15ex}{\scalebox{1.35}{$\cdot$}}}\,{\color{black!55}\raisebox{0.15ex}{\scalebox{1.35}{$\cdot$}}}\,{\color{black!55}\raisebox{0.15ex}{\scalebox{1.35}{$\cdot$}}}\,\\
    QR & 90 & $0$ & .259 & .349 & .363 & $.83$ & $0.35$ & $+1$ & $0.88$ & $0.16$ & --- & $2.47$ & $+20.8$ & \ding{55}\,\ding{51}\,{\color{black!55}\raisebox{0.15ex}{\scalebox{1.35}{$\cdot$}}}\,\ding{55}\,\ding{55}\,\\
    S5-120M & 120M & 500 & .098 & .163 & .133 & $2.32$ & $0.55$ & $-7$ & \cellcolor{TabHL}{\boldmath $0.98{\pm}.05$} & $7.36$ & $23$ & \cellcolor{TabHL}{\boldmath $0.99$} & $-0.7$ & \ding{55}\,\ding{51}\,\ding{51}\,\ding{55}\,\ding{55}\,\\
    S5-4k & 55M & 4k & .120 & .181 & .158 & $1.94$ & $0.54$ & $-4$ & \cellcolor{TabHL}{\boldmath $0.97{\pm}.05$} & $7.49$ & $23$ & \cellcolor{TabHL}{\boldmath $1.11$} & $-0.6$ & \ding{55}\,\ding{51}\,\ding{51}\,\ding{55}\,\ding{55}\,\\
    Mamba3 & 78M & 500 & .095 & .172 & .129 & $1.08$ & $0.59$ & $-24$ & \cellcolor{TabHL}{\boldmath $1.15{\pm}.04$} & $19.05$ & $7.7$ & \cellcolor{TabHL}{\boldmath $1.67$} & $+5.3$ & \ding{55}\,\ding{51}\,\ding{51}\,\ding{55}\,\ding{55}\,\\
    Mamba3-4k & 78M & 4k$^{\dagger}$ & .113 & .184 & .147 & $1.39$ & $0.62$ & $-27$ & \cellcolor{TabHL}{\boldmath $1.11{\pm}.05$} & $18.98$ & $9.4$ & \cellcolor{TabHL}{\boldmath $1.03$} & $-0.5$ & \ding{55}\,\ding{51}\,\ding{51}\,\ding{55}\,\ding{55}\,\\
    GDN & 94M & 500 & .093 & .181 & .152 & $1.99$ & $0.65$ & $-29$ & \cellcolor{TabHL}{\boldmath $1.12{\pm}.05$} & $15.43$ & $13$ & \cellcolor{TabHL}{\boldmath $0.89$} & $-1.5$ & \ding{55}\,\ding{51}\,\ding{51}\,\ding{55}\,\ding{55}\,\\
    \midrule
    \multicolumn{15}{@{}l}{\textbf{GOOG}} \\
    \emph{target} & --- & --- & low & low & low & $1.77$ & $0.54$ & ${\approx}0$ & {\boldmath$0.5$}\,{\tiny(exp)} & $0.5$\,{\tiny(exp)} & --- & {\boldmath$\tfrac{2}{3}$}\,{\tiny(exp)} & ${\sim}{-1.5}$\,{\tiny(exp)} & --- \\
    \midrule
    Historic & --- & --- & .069 & .133 & .118 & $1.77$ & $0.54$ & $0$ & ($-0.12$)$^{u}$ & ($0.01$)$^{u}$ & --- & ($0.00$)$^{u}$ & ($-0.1$)$^{u}$ & \ding{51}\,\ding{55}\,{\color{black!55}\raisebox{0.15ex}{\scalebox{1.35}{$\cdot$}}}\,{\color{black!55}\raisebox{0.15ex}{\scalebox{1.35}{$\cdot$}}}\,{\color{black!55}\raisebox{0.15ex}{\scalebox{1.35}{$\cdot$}}}\,\\
    Heuristic & 2$^{f}$ & $\infty$ & .069 & .133 & .118 & $1.77$ & $0.54$ & $0$ & $0.81$ & $0.73$ & --- & $1.00$ & $-0.1^{c}$ & \ding{51}\,\ding{51}\,{\color{black!55}\raisebox{0.15ex}{\scalebox{1.35}{$\cdot$}}}\,\ding{55}\,\ding{55}\,\\
    Propagator & 2$^{f}$ & $\infty$ & .069 & .133 & .118 & $1.77$ & $0.54$ & $0$ & ($0.73$)$^{u}$ & ($0.46$)$^{u}$ & --- & ($0.95$)$^{u}$ & ($-0.1$)$^{u}$ & \ding{51}\,\ding{55}\,{\color{black!55}\raisebox{0.15ex}{\scalebox{1.35}{$\cdot$}}}\,{\color{black!55}\raisebox{0.15ex}{\scalebox{1.35}{$\cdot$}}}\,{\color{black!55}\raisebox{0.15ex}{\scalebox{1.35}{$\cdot$}}}\,\\
    OW & 2$^{f}$ & exp. & .046 & .091 & .077 & $1.20$ & $0.54$ & $+1$ & ($0.72$)$^{u}$ & ($0.47$)$^{u}$ & --- & ($0.92$)$^{u}$ & ($-0.1$)$^{u}$ & \ding{55}\,\ding{55}\,{\color{black!55}\raisebox{0.15ex}{\scalebox{1.35}{$\cdot$}}}\,{\color{black!55}\raisebox{0.15ex}{\scalebox{1.35}{$\cdot$}}}\,{\color{black!55}\raisebox{0.15ex}{\scalebox{1.35}{$\cdot$}}}\,\\
    CST & 1.5k & $0$ & .316 & .501 & .613 & $-.44$ & $0.51$ & $+1$ & n.i. & n.i. & --- & ($1.84$)$^{u}$ & ($+4.5$)$^{u}$ & \ding{55}\,\ding{55}\,{\color{black!55}\raisebox{0.15ex}{\scalebox{1.35}{$\cdot$}}}\,{\color{black!55}\raisebox{0.15ex}{\scalebox{1.35}{$\cdot$}}}\,{\color{black!55}\raisebox{0.15ex}{\scalebox{1.35}{$\cdot$}}}\,\\
    NMZI & 1.5k$+$2 & EWMA & .318 & .503 & .612 & $.04$ & $0.25$ & $-1$ & n.i. & n.i. & --- & ($1.04$)$^{u}$ & ($+3.9$)$^{u}$ & \ding{55}\,\ding{55}\,{\color{black!55}\raisebox{0.15ex}{\scalebox{1.35}{$\cdot$}}}\,{\color{black!55}\raisebox{0.15ex}{\scalebox{1.35}{$\cdot$}}}\,{\color{black!55}\raisebox{0.15ex}{\scalebox{1.35}{$\cdot$}}}\,\\
    Hawkes & 0.4k & ${\lesssim}1$s & .408 & .461 & .636 & $1.91$ & $0.03$ & $0$ & ($0.76$)$^{u}$ & ($0.05$)$^{u}$ & --- & ($-2.01$)$^{u}$ & ($-34.6$)$^{u}$ & \ding{55}\,\ding{55}\,{\color{black!55}\raisebox{0.15ex}{\scalebox{1.35}{$\cdot$}}}\,{\color{black!55}\raisebox{0.15ex}{\scalebox{1.35}{$\cdot$}}}\,{\color{black!55}\raisebox{0.15ex}{\scalebox{1.35}{$\cdot$}}}\,\\
    QR & 90 & $0$ & .398 & .445 & .617 & $1.11$ & $0.36$ & $0$ & n.i. & n.i. & --- & $1.68$ & $+5.9$ & \ding{55}\,\ding{55}\,{\color{black!55}\raisebox{0.15ex}{\scalebox{1.35}{$\cdot$}}}\,{\color{black!55}\raisebox{0.15ex}{\scalebox{1.35}{$\cdot$}}}\,{\color{black!55}\raisebox{0.15ex}{\scalebox{1.35}{$\cdot$}}}\,\\
    S5-120M & 120M & 500 & .124 & .201 & .242 & $1.66$ & $0.56$ & $-18$ & \cellcolor{TabHL}{\boldmath $0.73{\pm}.06$} & $1.70$ & $26$ & \cellcolor{TabHL}{\boldmath $1.14$} & $+0.4$ & \ding{55}\,\ding{51}\,\ding{51}\,\ding{55}\,\ding{55}\,\\
    S5-4k & 55M & 4k & .131 & .217 & .232 & $1.92$ & $0.55$ & $+1$ & \cellcolor{TabHL}{\boldmath $0.75{\pm}.05$} & $1.83$ & $23$ & \cellcolor{TabHL}{\boldmath $1.16$} & $+0.9$ & \ding{55}\,\ding{51}\,\ding{51}\,\ding{55}\,\ding{55}\,\\
    Mamba3 & 78M & 500 & .117 & .224 & .124 & $1.95$ & $0.59$ & $-42$ & \cellcolor{TabHL}{\boldmath $0.97{\pm}.10$} & $6.24$ & $6.9$ & \cellcolor{TabHL}{\boldmath $1.67$} & $+2.6$ & \ding{55}\,\ding{51}\,\ding{51}\,\ding{55}\,\ding{55}\,\\
    Mamba3-4k & 78M & 4k$^{\dagger}$ & .133 & .233 & .181 & $.97$ & $0.59$ & $-879$ & \cellcolor{TabHL}{\boldmath $1.21{\pm}.11$} & $12.56$ & $3.4$ & \cellcolor{TabHL}{\boldmath $0.24$} & $-13.8$ & \ding{55}\,\ding{51}\,\ding{51}\,\ding{55}\,\ding{55}\,\\
    GDN & 94M & 500 & .134 & .222 & .263 & $.79$ & $0.54$ & $-85$ & \cellcolor{TabHL}{\boldmath $1.11{\pm}.10$} & $8.42$ & $9.6$ & \cellcolor{TabHL}{\boldmath $0.18$} & $-8.5$ & \ding{55}\,\ding{51}\,\ding{51}\,\ding{55}\,\ding{55}\,\\
    \midrule
    \multicolumn{15}{@{}l}{\textbf{EA}$^{\ddagger}$} \\
    \emph{target} & --- & --- & low & low & low & $1.31$ & $0.47$ & ${\approx}0$ & {\boldmath$0.5$}\,{\tiny(exp)} & $0.5$\,{\tiny(exp)} & --- & {\boldmath$\tfrac{2}{3}$}\,{\tiny(exp)} & ${\sim}{-1.5}$\,{\tiny(exp)} & --- \\
    \midrule
    Historic & --- & --- & .049 & .113 & .112 & $1.31$ & $0.47$ & $0$ & ($-0.52$)$^{u}$ & ($0.00$)$^{u}$ & --- & ($0.00$)$^{u}$ & ($-0.5$)$^{u}$ & \ding{51}\,\ding{55}\,{\color{black!55}\raisebox{0.15ex}{\scalebox{1.35}{$\cdot$}}}\,{\color{black!55}\raisebox{0.15ex}{\scalebox{1.35}{$\cdot$}}}\,{\color{black!55}\raisebox{0.15ex}{\scalebox{1.35}{$\cdot$}}}\,\\
    Heuristic & 2$^{f}$ & $\infty$ & .049 & .113 & .112 & $1.31$ & $0.47$ & $0$ & ($0.85$)$^{u}$ & ($1.42$)$^{u}$ & --- & ($1.01$)$^{u}$ & ($-0.1$)$^{u}$ & \ding{51}\,\ding{55}\,{\color{black!55}\raisebox{0.15ex}{\scalebox{1.35}{$\cdot$}}}\,{\color{black!55}\raisebox{0.15ex}{\scalebox{1.35}{$\cdot$}}}\,{\color{black!55}\raisebox{0.15ex}{\scalebox{1.35}{$\cdot$}}}\,\\
    Propagator & 2$^{f}$ & $\infty$ & .049 & .113 & .112 & $1.31$ & $0.47$ & $0$ & ($0.76$)$^{u}$ & ($0.93$)$^{u}$ & --- & ($0.97$)$^{u}$ & ($-0.2$)$^{u}$ & \ding{51}\,\ding{55}\,{\color{black!55}\raisebox{0.15ex}{\scalebox{1.35}{$\cdot$}}}\,{\color{black!55}\raisebox{0.15ex}{\scalebox{1.35}{$\cdot$}}}\,{\color{black!55}\raisebox{0.15ex}{\scalebox{1.35}{$\cdot$}}}\,\\
    OW & 2$^{f}$ & exp. & .045 & .097 & .099 & $1.25$ & $0.43$ & $0$ & ($0.74$)$^{u}$ & ($0.97$)$^{u}$ & --- & ($0.82$)$^{u}$ & ($-0.5$)$^{u}$ & \ding{55}\,\ding{55}\,{\color{black!55}\raisebox{0.15ex}{\scalebox{1.35}{$\cdot$}}}\,{\color{black!55}\raisebox{0.15ex}{\scalebox{1.35}{$\cdot$}}}\,{\color{black!55}\raisebox{0.15ex}{\scalebox{1.35}{$\cdot$}}}\,\\
    CST & 1.5k & $0$ & .211 & .356 & .372 & $.37$ & $0.44$ & $+1$ & n.i. & n.i. & --- & ($3.49$)$^{u}$ & ($+16.6$)$^{u}$ & \ding{55}\,\ding{55}\,{\color{black!55}\raisebox{0.15ex}{\scalebox{1.35}{$\cdot$}}}\,{\color{black!55}\raisebox{0.15ex}{\scalebox{1.35}{$\cdot$}}}\,{\color{black!55}\raisebox{0.15ex}{\scalebox{1.35}{$\cdot$}}}\,\\
    NMZI & 1.5k$+$2 & EWMA & .214 & .358 & .372 & $1.02$ & $0.45$ & $0$ & n.i. & n.i. & --- & ($5.43$)$^{u}$ & ($+16.6$)$^{u}$ & \ding{55}\,\ding{55}\,{\color{black!55}\raisebox{0.15ex}{\scalebox{1.35}{$\cdot$}}}\,{\color{black!55}\raisebox{0.15ex}{\scalebox{1.35}{$\cdot$}}}\,{\color{black!55}\raisebox{0.15ex}{\scalebox{1.35}{$\cdot$}}}\,\\
    Hawkes & 0.4k & ${\lesssim}1$s & .312 & .382 & .431 & $2.08$ & $0.43$ & $0$ & ($0.45$)$^{u}$ & ($0.31$)$^{u}$ & --- & ($0.80$)$^{u}$ & ($+0.7$)$^{u}$ & \ding{55}\,\ding{55}\,{\color{black!55}\raisebox{0.15ex}{\scalebox{1.35}{$\cdot$}}}\,{\color{black!55}\raisebox{0.15ex}{\scalebox{1.35}{$\cdot$}}}\,{\color{black!55}\raisebox{0.15ex}{\scalebox{1.35}{$\cdot$}}}\,\\
    QR & 90 & $0$ & .302 & .374 & .419 & $.23$ & $0.45$ & $0$ & $0.74$ & $0.32$ & --- & $0.42$ & $-1.0$ & \ding{55}\,\ding{51}\,{\color{black!55}\raisebox{0.15ex}{\scalebox{1.35}{$\cdot$}}}\,\ding{55}\,\ding{51}\,\\
    S5-120M & 120M & 500 & .135 & .221 & .224 & $1.68$ & $0.57$ & $-3$ & \cellcolor{TabHL}{\boldmath $0.84{\pm}.06$} & $12.23$ & $29$ & \cellcolor{TabHL}{\boldmath $1.18$} & n.i. & \ding{55}\,\ding{51}\,\ding{51}\,\ding{55}\,\ding{55}\,\\
    S5-4k & 55M & 4k & .145 & .264 & .275 & $1.62$ & $0.56$ & $+4$ & \cellcolor{TabHL}{\boldmath $1.05{\pm}.06$} & $19.18$ & $38$ & \cellcolor{TabHL}{\boldmath $1.24$} & $+0.6$ & \ding{55}\,\ding{51}\,\ding{51}\,\ding{55}\,\ding{55}\,\\
    Mamba3 & 78M & 500 & .125 & .182 & .203 & $2.03$ & $0.59$ & $-4$ & \cellcolor{TabHL}{\boldmath $1.10{\pm}.07$} & $49.52$ & $25$ & \cellcolor{TabHL}{\boldmath $1.37$} & $+2.2$ & \ding{55}\,\ding{51}\,\ding{51}\,\ding{55}\,\ding{55}\,\\
    Mamba3-4k & 78M & 4k$^{\dagger}$ & .113 & .179 & .210 & $1.82$ & $0.62$ & $-13$ & \cellcolor{TabHL}{\boldmath $1.12{\pm}.07$} & $41.66$ & $20$ & \cellcolor{TabHL}{\boldmath $1.14$} & $+0.1$ & \ding{55}\,\ding{51}\,\ding{51}\,\ding{55}\,\ding{55}\,\\
    GDN & 94M & 500 & .137 & .172 & .202 & $2.11$ & $0.57$ & $-6$ & \cellcolor{TabHL}{\boldmath $1.06{\pm}.07$} & $43.54$ & $13$ & \cellcolor{TabHL}{\boldmath $1.34$} & $+1.6$ & \ding{55}\,\ding{51}\,\ding{51}\,\ding{55}\,\ding{55}\,\\
    \midrule
    \multicolumn{15}{@{}l}{\textbf{MSFT}} \\
    \emph{target} & --- & --- & low & low & low & $1.51$ & $0.55$ & ${\approx}0$ & {\boldmath$0.5$}\,{\tiny(exp)} & $0.5$\,{\tiny(exp)} & --- & {\boldmath$\tfrac{2}{3}$}\,{\tiny(exp)} & ${\sim}{-1.5}$\,{\tiny(exp)} & --- \\
    \midrule
    Historic & --- & --- & .066 & .149 & .105 & $1.51$ & $0.55$ & $+12$ & ($0.87$)$^{u}$ & ($0.23$)$^{u}$ & --- & ($0.01$)$^{u}$ & ($-0.9$)$^{u}$ & \ding{51}\,\ding{55}\,{\color{black!55}\raisebox{0.15ex}{\scalebox{1.35}{$\cdot$}}}\,{\color{black!55}\raisebox{0.15ex}{\scalebox{1.35}{$\cdot$}}}\,{\color{black!55}\raisebox{0.15ex}{\scalebox{1.35}{$\cdot$}}}\,\\
    Heuristic & 2$^{f}$ & $\infty$ & .066 & .149 & .105 & $1.51$ & $0.55$ & $+12$ & $0.84$ & $1.29$ & --- & $0.89$ & $-0.3^{c}$ & \ding{51}\,\ding{51}\,{\color{black!55}\raisebox{0.15ex}{\scalebox{1.35}{$\cdot$}}}\,\ding{55}\,\ding{55}\,\\
    Propagator & 2$^{f}$ & $\infty$ & .066 & .149 & .105 & $1.51$ & $0.55$ & $+12$ & $0.83$ & $0.94$ & --- & $0.82$ & $-0.5$ & \ding{51}\,\ding{51}\,{\color{black!55}\raisebox{0.15ex}{\scalebox{1.35}{$\cdot$}}}\,\ding{55}\,\ding{55}\,\\
    OW & 2$^{f}$ & exp. & .050 & .100 & .069 & $1.82$ & $0.55$ & $-7$ & $0.80$ & n.i. & --- & $0.71$ & $-0.5$ & \ding{55}\,\ding{51}\,{\color{black!55}\raisebox{0.15ex}{\scalebox{1.35}{$\cdot$}}}\,{\color{black!55}\raisebox{0.15ex}{\scalebox{1.35}{$\cdot$}}}\,{\color{black!55}\raisebox{0.15ex}{\scalebox{1.35}{$\cdot$}}}\,\\
    CST & 1.5k & $0$ & .339 & .462 & .590 & $.20$ & $0.24$ & $-13$ & n.i. & n.i. & --- & ($1.75$)$^{u}$ & ($+5.1$)$^{u}$ & \ding{55}\,\ding{55}\,{\color{black!55}\raisebox{0.15ex}{\scalebox{1.35}{$\cdot$}}}\,{\color{black!55}\raisebox{0.15ex}{\scalebox{1.35}{$\cdot$}}}\,{\color{black!55}\raisebox{0.15ex}{\scalebox{1.35}{$\cdot$}}}\,\\
    NMZI & 1.5k$+$2 & EWMA & .346 & .476 & .597 & n.i. & $0.17$ & $0$ & n.i. & n.i. & --- & ($-2.01$)$^{u}$ & ($-37.8$)$^{u}$ & {\color{black!55}\raisebox{0.15ex}{\scalebox{1.35}{$\cdot$}}}\,\ding{55}\,{\color{black!55}\raisebox{0.15ex}{\scalebox{1.35}{$\cdot$}}}\,{\color{black!55}\raisebox{0.15ex}{\scalebox{1.35}{$\cdot$}}}\,{\color{black!55}\raisebox{0.15ex}{\scalebox{1.35}{$\cdot$}}}\,\\
    Hawkes & 0.4k & ${\lesssim}1$s & .384 & .446 & .599 & $1.88$ & $0.23$ & $+2$ & n.i. & n.i. & --- & $0.57$ & $-6.4$ & \ding{55}\,\ding{55}\,{\color{black!55}\raisebox{0.15ex}{\scalebox{1.35}{$\cdot$}}}\,{\color{black!55}\raisebox{0.15ex}{\scalebox{1.35}{$\cdot$}}}\,{\color{black!55}\raisebox{0.15ex}{\scalebox{1.35}{$\cdot$}}}\,\\
    QR & 90 & $0$ & .369 & .429 & .582 & $.88$ & $0.50$ & $-1$ & $0.71$ & n.i. & --- & $1.44$ & $+4.6$ & \ding{55}\,\ding{51}\,{\color{black!55}\raisebox{0.15ex}{\scalebox{1.35}{$\cdot$}}}\,{\color{black!55}\raisebox{0.15ex}{\scalebox{1.35}{$\cdot$}}}\,\ding{55}\,\\
    S5-120M & 120M & 500 & .214 & .283 & .325 & $1.96$ & $0.55$ & $-30$ & \cellcolor{TabHL}{\boldmath $0.63{\pm}.11$} & $3.07$ & $11$ & \cellcolor{TabHL}{\boldmath $1.14$} & $+0.3$ & \ding{55}\,\ding{51}\,\ding{51}\,\ding{55}\,\ding{55}\,\\
    S5-4k & 55M & 4k & .226 & .298 & .277 & $1.39$ & $0.59$ & $+18$ & \cellcolor{TabHL}{\boldmath $0.69{\pm}.11$} & $3.48$ & $18$ & \cellcolor{TabHL}{\boldmath $1.21$} & $+0.5$ & \ding{55}\,\ding{51}\,\ding{51}\,\ding{55}\,\ding{55}\,\\
    Mamba3 & 78M & 500 & .163 & .282 & .208 & $1.47$ & $0.62$ & $-39$ & \cellcolor{TabHL}{\boldmath $0.81{\pm}.11$} & $7.60$ & $8.8$ & \cellcolor{TabHL}{\boldmath $1.19$} & $+0.2$ & \ding{55}\,\ding{51}\,\ding{51}\,\ding{55}\,\ding{55}\,\\
    Mamba3-4k & 78M & 4k$^{\dagger}$ & .134 & .278 & .160 & $1.82$ & $0.60$ & $-62$ & \cellcolor{TabHL}{\boldmath $0.86{\pm}.14$} & $8.79$ & $9.6$ & \cellcolor{TabHL}{\boldmath $0.72$} & $-3.9$ & \ding{55}\,\ding{51}\,\ding{51}\,\ding{55}\,\ding{55}\,\\
    GDN & 94M & 500 & .178 & .276 & .213 & $1.03$ & $0.52$ & $-58$ & \cellcolor{TabHL}{\boldmath $0.79{\pm}.12$} & $8.28$ & $11$ & \cellcolor{TabHL}{\boldmath $0.95$} & $-2.1$ & \ding{55}\,\ding{51}\,\ding{51}\,\ding{55}\,\ding{55}\,\\
    \bottomrule
  \end{tabular}
\end{table*}

\subsection{Phase 2: Metaorder response}
\label{sec:res-beta}

The middle band of Figure~\ref{fig:midtraj} shows the build-up. We first check that a law exists, then measure it. All five neural generators show a sharp minimum of the $L_2$ misfit over $\delta$ (Section~\ref{sec:method}), so their exponent is identified: claim C2 passes across the panel. For the Historic replay the misfit is flat and the amplitude sits at zero: with no signal, no $\delta$ fits better than another, so there is no law, the correct answer for a stream that cannot react.

Where a law is identified, we calculate $\delta$. On AMD the binned exponents are $0.66$ (S5-120M), $0.79$ (S5-4k), $0.87$ (Mamba3), $0.90$ (GDN), and $0.93$ (Mamba3-4k). Their $95\%$ day-clustered bootstrap intervals are $\pm 0.05$--$0.06$: the S5 family stays closest to $0.5$, yet no interval comes near it. The band is stable under every estimator variant of Section~\ref{sec:method}, and the estimator's bias points down: the true build-up is, if anything, steeper.

The amplitude turns the overshoot into money: the neural panel shows $Y = 2.2$--$8.6$ against the empirical $Y \approx 0.5$, four to seventeen times what a real metaorder pays per unit of participation. In levels, a $100$-child metaorder moves the simulated price by $17$--$70$\,bps, against at most $11$\,bps for any baseline (the Heuristic replay). On AMD, claim C4 fails in every neural cell: no exponent within $0.10$ of $0.5$, no amplitude inside the empirical band.

The baselines behave as designed. The three replay kernels (Heuristic, Propagator, OW) are identified at $\delta = 0.73$--$0.81$ with $Y = 0.96$--$1.46$: the mechanical response written into them, at roughly the price a real book charges. Their $\delta$ sits below the closed-form values (${\approx}0.9$--$1.0$) by the small-amplitude bias measured in the planted-law calibration. QR is identified at $\delta = 0.72$ with $Y = 0.09$, a law with almost no impact behind it; the zero-intelligence pair (CST, NMZI) and Hawkes identify nothing at all ($Y = 0.03$ for Hawkes on AMD, below the floor). The neural models differ from the baselines in amplitude: their exponents overlap with the replay kernels', but their $Y$ exceeds every baseline's. The blind control shows that this impact is the models' own (C3): the visible peak is at least $3\times$ the blind floor in all $25$ neural cells and over $10\times$ in $16$ (v/b, Table~\ref{tab:scorecard}). This excess is the indirect impact the experiments were designed to isolate.

\subsection{Phase 3: Relaxation}
\label{sec:res-relax}

The bottom band asks whether impact comes down once execution stops. Real metaorders fall to about two-thirds of peak within one to two execution durations and to about one-half within ten~\citep{moro2009market, bucci2019slow, bouchaud2018trades}, the span of our window, so a flat curve cannot be blamed on it. The kernel baselines set the scale: at realistic child spacing, a kernel spends most of its transient between children, so its closed-form end-over-peak is $0.94$ for the Propagator and $0.87$ for OW, and the measured values land beside these ($0.89$ and $0.81$ on AMD, against the permanent Heuristic's $0.95$).

The neural panel shows no credible relaxation. End-over-peak on AMD reads $1.19$ (S5-120M), $1.15$ (S5-4k), $1.16$ (Mamba3), $1.36$ (Mamba3-4k), and $0.88$ (GDN), with post-execution slopes of $+0.6$, $\pm0.0$, $+0.4$, $+3.7$, and $-2.4\%$ of the peak per execution duration. Four of the five remain above peak: the price keeps moving in the metaorder's direction after the last child has executed. GDN falls below peak, but its free drift of $-127$ ticks prevents attributing the decline to resilience. That is claim C5, and no neural cell passes it.

\subsection{Extending to a larger universe}
\label{sec:res-panel}

The AMD portrait replicates down the blocks of Table~\ref{tab:scorecard}: the neural panel is one claim-ladder signature repeated $25$ times: C1 fails, C2 and C3 pass, C4 and C5 fail.

All five neural generators identify a law on every stock, every exponent above the square-root reference: $0.63$--$0.93$ on AMD and MSFT, $0.73$--$1.21$ on GOOG, $0.97$--$1.15$ on NVDA, $0.84$--$1.12$ on EA. All $25$ day-clustered intervals (the $\pm$ of Table~\ref{tab:scorecard}) exclude $0.5$, with half-widths of $0.04$--$0.14$ and the tightest case (MSFT S5-120M) still bounding its exponent above $0.53$. The replay returns no law on any stock, keeping the estimator honest. Ranking the generators by $\delta$ or by $Y$ gives nearly the same order on every stock (Kendall $W = 0.82$ and $0.81$): the benchmark measures a model-specific component that persists across stocks.

In $20$ of the $25$ neural rows, $H$ exceeds the stock's own value; in three it matches, and only GDN, on AMD and MSFT, sits below. No neural row reaches two thirds by decay: the three lowest end-over-peak entries (GDN $0.18$ and Mamba3-4k $0.24$ on GOOG; Mamba3-4k $0.72$ on MSFT) sit on the heaviest-drift rows of their blocks; the descent is drift, not resilience.

On the four stocks in the training corpus, $Y \approx 2$--$9$ on AMD and MSFT, $1.7$--$12.6$ on GOOG and $7$--$19$ on NVDA. On EA, the one stock outside the corpus, $Y = 12$--$50$. The observed OOD result does not destroy the law; it destroys its calibration.

The Propagator and OW are identified on AMD, NVDA and MSFT but not on GOOG or EA; QR's end-over-peak swings from $2.47$ (NVDA) to $0.42$ (EA).

The three $500$-context generators sit close together in the distributional columns yet differ materially in impact. Across stocks, the correlation between the LOB-Bench KS ranking and the $|\delta - 0.5|$ ranking is unstable ($+0.54$ on GOOG against $-0.71$ on MSFT, $p = 0.02$) and mostly negative within the neural family (MSFT neural rows: $-0.90$, $p = 0.08$). The better mimic can sit farther from the impact law, so the two properties must be measured separately.

\section{Discussion}
\label{sec:discussion}

Our results separate three questions that distributional evaluation conflates: whether impact exists, whether it is the generator's own, and whether its dynamics are market-realistic; across the neural panel, the answer is yes, yes, and no. The joint C4/C5 failure is consistent with insufficient replenishment: in a real book, each child's push decays before the next arrives, turning $N$ pushes into $\sqrt{N}$ rather than $N$. With persistent flow but insufficient replenishment, impact instead builds up too steeply and fails to relax credibly (Sections~\ref{sec:res-flow}--\ref{sec:res-relax}). The blind control shows that the build-up is not explained by book mechanics alone; the defect hides at the single-event scale. We attribute the missing replenishment to limited long-horizon memory: relaxation requires retaining a reference price from before the metaorder, which the models' still-short context windows do not preserve.

These impact errors matter directly for execution: an execution cost model depends on the shape, amplitude, and relaxation of market impact~\citep{gueant2016financial, mohl2025jaxmarlhft}, all measured here per model and stock. Today's generators may achieve message-level realism, but they remain unsuitable as environments for execution agents while the endogenous impact they do produce has the wrong shape and does not decay: such a simulator can reward schedules that a real book would punish.

The results suggest how to improve these models. At each training checkpoint, Phase~1 can be run on freely generated rollouts: the goal is for the generated $\gamma_{\mathrm{flow}}$ and $H$ to approach their empirical values for the stock. The direction matters: learn replenishment rather than unlearn persistence; killing flow memory would restore the balance vacuously. Phases~2 and~3 then test whether the repair restores realistic build-up and relaxation. The 4k-context variants fail the same tests but remain well short of a metaorder, leaving the attribution untested rather than refuted; longer contexts and larger models are the natural next test.

\paragraph{Limitations.} One stock is reported band by band, the other four only through Table~\ref{tab:scorecard}, and everything is measured at one operating point: one participation rate, one child size, aggressive children~\citep{sato2025universality, maitrier2025double}, one metaorder shape, one month of one exchange; and $\delta$ depends on that point. The $10\%$ target lies inside the retained empirical samples~\citep{bacry2015market, zarinelli2015beyond}; the extreme-volume regime~\citep{bucci2019crossover} is not probed. The month itself is a sample: a month atypical by about one within-month standard deviation would already move the exponent past its half-width cap.

One open-source package (repository \url{https://github.com/george-nigmatulin/lob-impact}) contains the protocol, calibration, estimators, controls, and reporting rules; a user supplies a model and one month of LOBSTER-format data for any stock.

\section{Conclusion}
\label{sec:conclusion}

Market impact emerges in generative LOB models. Five neural message generators, trained only on next-token prediction, answer an injected metaorder with an identifiable law in all 25 model–stock pairs, while a replay that cannot react does not answer at all. The models fail in two related ways: impact builds up more steeply than the square-root reference and shows no credible post-execution relaxation. Together, these failures point to insufficient replenishment to compensate persistent order flow. We tentatively attribute both to limited long-horizon memory, induced by still-short context windows, a concrete target for future models.

\begin{acks}
This work was supported by the UKRI ESRC Grand Union DTP [ES/Y001761/1]; EPSRC [UKRI5149] and other \href{https://www.bold-lab.ai}{sponsors of BOLD}; the Department of Computer Science, University of Oxford; and a UKRI AI World Leading Researcher Fellowship [EP/W002949/1].
The authors acknowledge the use of resources provided by the Isambard-AI National AI Research Resource (AIRR)~\cite{mcintoshsmith2024isambardaileadershipclasssupercomputer}. Isambard-AI is operated by the University of Bristol and is funded by the UK Government’s Department for Business, Innovation, Science and Trade (DBIST) via UK Research and Innovation; and the Science and Technology Facilities Council [ST/AIRR/I-A-I/1023].
\end{acks}
\bibliographystyle{ACM-Reference-Format}
\bibliography{sample-base}

\end{document}